\documentclass{aa}
\usepackage{graphicx}
\usepackage{subcaption}
\usepackage{amssymb}
\usepackage{amsmath}
\usepackage{float}
\usepackage{txfonts}
\usepackage{gensymb}
\usepackage{hyperref}
\usepackage[normalem]{ulem}
\usepackage[figuresright]{rotating}

\begin{document}

\authorrunning{Qingshun Hu}
\titlerunning{Decoding Clusters' Morphological Evolution}

\title{Decoding the morphological evolution of open clusters}

\author{Qingshun Hu \inst{1,2} \and Yu Zhang \inst{1,2} \and Ali Esamdin \inst{1,2}}

\institute{Xinjiang Astronomical Observatory, Chinese Academy of Sciences, Urumqi, Xinjiang 830011, People's Republic of China, (\email{zhy@xao.ac.cn}, \email{aliyi@xao.ac.cn} \label{inst1}) \and University of Chinese Academy of Sciences, Beijing 100049, People's Republic of China\label{inst2}
}

\date{Received 2021 June 3; accepted 2021 September 8}

\abstract{The properties of open clusters such as metallicity, age, and morphology are useful tools in studies of the dynamic evolution of open clusters. The morphology of open clusters can help us better understand the evolution of such structures.}
{We aim to analyze the morphological evolution of 1256 open clusters by combining the shapes of the sample clusters in the proper motion space with their morphology in the two-dimensional spherical Galactic coordinate system, providing their shape parameters based on a member catalog derived from {\it Gaia} Second Data Release as well as data from the literature.}
{We applied a combination of a nonparametric bivariate density estimation with the least square ellipse fitting  to derive the shape parameters of the sample clusters.}
{We derived the shape parameters of the sample clusters in the two-dimensional spherical Galactic coordinate system and that of the proper motion space. By analyzing the dislocation of the sample clusters, we find that the dislocation, $d,$ is related to the X-axis pointing toward the Galactic center, Y-axis pointing in the direction of Galactic rotation, and the Z-axis (log(|H|/pc)) that is positive toward the Galactic north pole. This finding underlines the important role of the dislocation of clusters in tracking the external environment of the Milky Way. The orientation ($q_{pm}$) of the clusters, with $e_{pm}$~$\geq$~0.4, presents an aggregate distribution in the range of -45$\degr$ to 45$\degr$ , comprising about 74\% of them. This probably suggests that these clusters tend to deform heavily in the direction of the Galactic plane.\ NGC~752 is in a slight stage of expansion in the two-dimensional space and will become deformed, in terms of its morphology, along the direction perpendicular to the original stretching direction in the future if no other events occur. The relative degree of deformation of the sample clusters in the short-axis direction decreases as their ages increase. On average, the severely distorted sample clusters in each group account for about 26\%~$\pm$~9\%. This possibly implies a uniform external environment in the range of $|$H$|$~$\leq$~300~pc if the sample completeness of each group is not taken into account.}
{}

\keywords{Methods: statistical - Galaxy: open clusters and associations - catalogs - Galaxy: stellar content
}

\maketitle{}

\section{Introduction}

With ages ranging from millions to billions of years and typically containing $10^2$-$10^4$ stars,  open clusters are irregular, dynamically bound stellar systems located in the Galactic disk. The properties of open cluster populations, such as metallicity, age, and morphology, are useful tools in the study of their dynamic evolution \citep{krum19}. Although the member stars of cluster populations are difficult to detect in full, particularly with regard to fainter members, a great deal of effort has been made to compile a large catalog of members of open clusters \citep[e.g.,][]{samp17, cant18, cast19, cast20, cant20a, cant20b} that would serve as a basis for statistical surveys of the shapes of open clusters.

The morphology of open clusters can mostly be viewed as a layered structure, namely, one containing a core and corona, as confirmed by \citet[e.g.,][]{khol69, rabo98b, chen04}. This means that the shapes of the coronas and central cores of open clusters dominated by different forces evolve in different ways. As an open cluster evolves, the distribution of member stars in its core is altered by internal gravitational interactions among its member stars. Subsequently, stellar evaporation and external disturbances, for instance, Galactic tidal force, differential rotation, and encounters with molecular clouds, would affect the external spatial structure of the cluster and eventually dissolve it. In light of the layered structure of open clusters, \citet{zhai17} studied the morphology of 154 open clusters selected from the WEBDA database. But these authors did not quantify the layered structure of the cluster shapes. In our latest paper in this series (\citet{hu21}, hereafter, Paper ~I), we reported the morphology of 265 open clusters, initially defined the radial and tangential stratification, and provided relevant conclusions. We also provided the shape parameters of the sample clusters. Nevertheless, due to the relatively small number of clusters in the sample studied in Paper~I, such conclusions are still insufficient for a quantitative investigation of the layered structure. Certain conclusions about clusters' evolution also require verification.

The studies mentioned above were implemented in two-dimensional space. Similarly, the morphology of our current sample clusters is also studied in two-dimensional space, whereas the true shape of star clusters is three-dimensional. As a consequence, we are confronted with the projection effect remaining in the present work. The degree of possible discrepancy depends on the axial relationship of the ellipsoid and the axial orientation relative to the cluster's line of sight \citep{pisk08}. The true spatial three-dimensional shape of most observed clusters is not yet known, especially for distant clusters. \citet{wiel74} predicted that the ratio of the three orthogonal axes of star clusters, considered as a tridimensional ellipsoid, should be 2.0:1.4:1.0. The major axis points to the galactic center and the minor axis is perpendicular to the galactic disk. At present, we are not able to study the three-dimensional morphology of a large sample observationally because the distance precision of clusters' members is not yet high enough, especially for distant single sources. While a study of the three-dimensional morphology of clusters was performed recently by \citet{pang21}, they only investigated the three-dimensional structure within the tidal radius of 13 open clusters near the Sun. This is still not enough to examine the clusters' layered structure. Thus, the statistical survey of morphology for a large sample in two dimensions is still meaningful and valuable.

In this work, we expand the sample clusters studied in Paper~I and modify and refine the definitions of the radial and tangential stratification degree of star clusters in Paper~I. And they are then used to quantitatively investigate the layered structure of star clusters and diagnose the morphological evolution of clusters. Meanwhile, we define, for the first time, a new physical quantity, namely, the morphological dislocation of open clusters. This enables us to measure the intensity of cluster morphological evolution and also to trace the complexity of the external environment. Furthermore, since the distribution of the members of open clusters in proper motion space contributes more or less to the overall morphology of the clusters in the future, the shape of open clusters may be related to the distribution of their members in the proper motion space. Also, for the first time, we statistically study the morphology of clusters in the proper motion space, seeking some possible correlations between the morphology in the proper motion space and in the two-dimensional spherical Galactic coordinate system. In addition, the shape parameters of the clusters in the Galactic spherical coordinate system are provided, as well as the shape parameters of the clusters in the proper motion space.

In the present paper, we report on the morphological evolution of open clusters through a statistical study of their layered structure and their morphology in proper motion space. We present, in Sect.~2, the data used in our work and the methods used to derive the shapes of the clusters in the Galactic spherical coordinate system (i.e., the overall versus the core) and to derive their morphology in the proper motion space. In Sect.~3, we present a set of results derived from our statistical analysis of the shape parameters of the clusters. Finally, in Sect.~4, we give a concise summary of our results.

\section{Data and methods}

We begin with the member catalog of open clusters \citep{cant20a}, which provides the combined spatiokinematic-photometric membership of 1481 open clusters based on the UPMASK method \citep{kron14} applied to {\it Gaia} Data Release~2 (DR2) data. Although \citet{cant20b} published the member catalog of 2017 open clusters, that is, the updated of the 1481 clusters from \citet{cant20a}, they only released members with probabilities ranging from 0.7 to 1; however, \citet{cant20a} provided all members with probability from 0.1 to 1 in their work. We sought to select a sample with age parameters and members covering all probabilities, thus, we adopted the age parameters of 2017 open clusters from \citet{cant20b} given that there  are  no age parameters given in \citet{cant20a}. After cross-matching the 1481 open clusters given by \citep{cant20a} with the 2017 clusters with the age parameters, we obtained 1256 clusters in common. These clusters offer the advantage of spanning a wide range of ages, from a few million years (young system) to several gigayears (old system), which meets the requirements of our study.

The method we adopt in the present paper is the same as employed in Paper~I and we extend it to the proper motion space. The method consists of a nonparametric bivariate density estimation and the least square ellipse fitting. In this way, we can delineate the shape of sample clusters in the two-dimensional (2D) spherical Galactic coordinate system and their morphology in the proper motion space. Finally, all shape parameters were derived to form the basis for our study of the morphological evolution of clusters.

\begin{figure*}[ht]
\centering
\includegraphics[width=120mm]{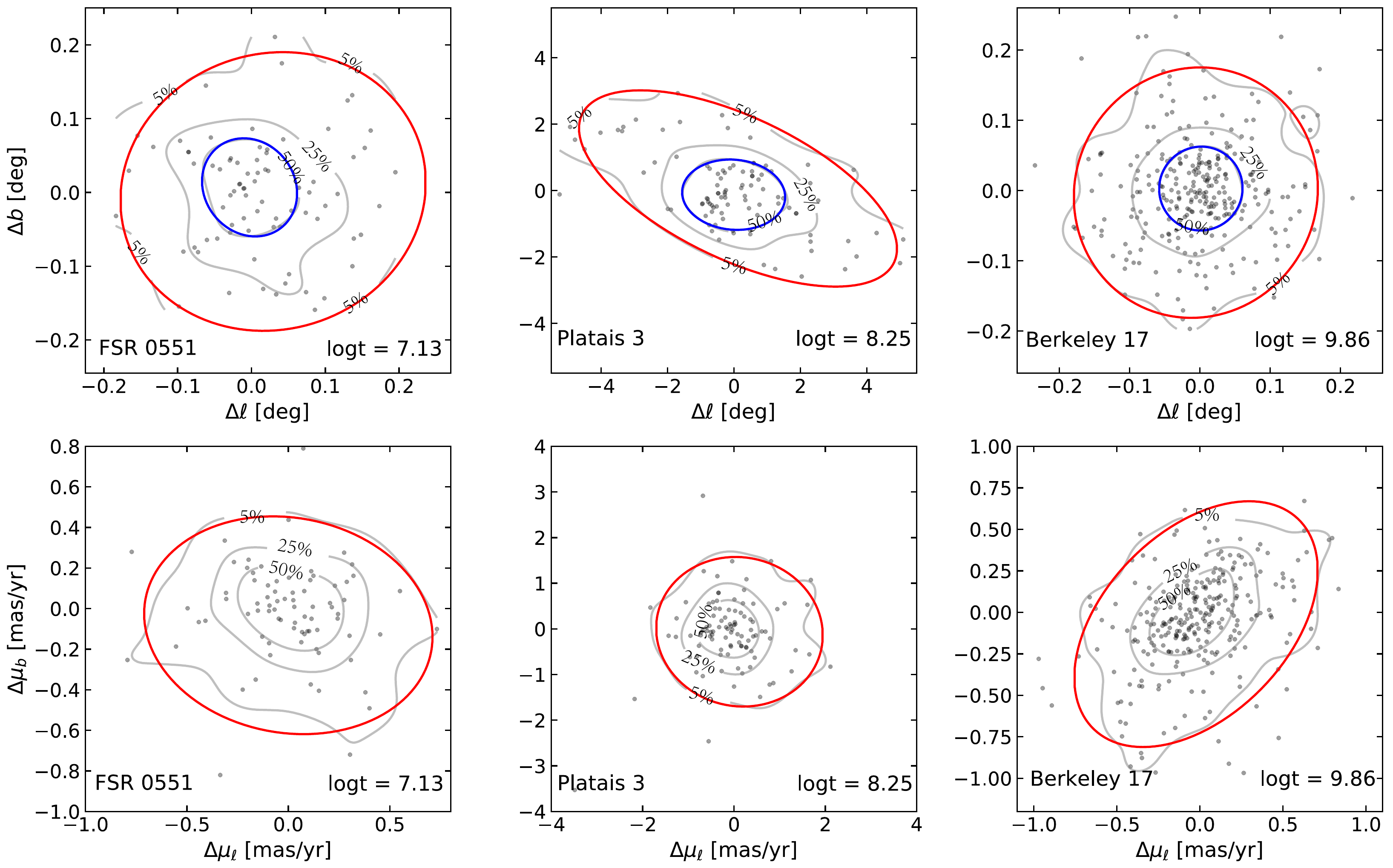}
\caption{
General scheme of fitting cluster density profiles in the 2D spherical Galactic coordinate system, as well as that in the proper motion space, for three open clusters (\ FSR~0551,\ Platais~3, and\ Berkeley~17). For each cluster in the top three panels, the black dots represent the member stars of a cluster with the gray lines as the stellar number density profiles at 50\% of and 5\% of the peak. The fitted ellipses of 50\% and 5\% of the peak stellar number density of the clusters are marked in blue and red, respectively. The percentage labels on the gray lines represent the relative percentage of the actual density value relative to the peak. For each cluster in the bottom three panels, the black dots still represent the member stars of a cluster with the gray lines as the stellar number density profiles at 5\% of the peak. The fitted ellipse of 5\% of the peak stellar number density of the clusters is marked in red. The density contours at 5\% in some panels are incomplete because a grid with specific boundaries was selected for the density evaluation.
}
\label{fig:Density_profile}
\end{figure*}

The process of the approach is as follows: the first step is to estimate the kernel density of the distribution of the cluster members in the Galactic coordinate system and the proper motion space, and then obtain the different iso-density contours of their distribution. The second step is to perform ellipse fitting for different equivalent density profiles to obtain the corresponding shape parameters. We note that we still assign two ellipses in the two-dimensional spherical Galactic coordinate system to represent the shape of the core of a cluster and the overall cluster. For this, we set the peak stellar number density of each open cluster in the two-dimensional system arbitrarily at 50\% and 5\%, respectively. But we only assign one ellipse to represent the morphology of clusters in the proper motion space, which is arbitrarily assigned to fit the iso-density line at 5\% of the peak stellar number density in the proper motion space. The result of this procedure is visualized in Figure~\ref{fig:Density_profile}, which displays three sample clusters (\ FSR~0551,\ Platais~3, and\ Berkeley~17).

The shape parameters we derived are as follows: in the two-dimensional spherical Galactic coordinate system, the shape parameters contain the ellipticities ($e_{core}$ denoting the ellipticity of the inner ellipse and $e_{all}$ the ellipticity of the overall ellipse), the orientations ($q_{core}$ representing the orientation of the inner ellipse and $q_{all}$ the orientation of the overall ellipse), and some additional parameters ($a_{core}$ referring to half-length axes of inner ellipses, $a_{all}$ half-length axes of overall ellipse, $c_{core}$ half-short axes of inner ellipse, $c_{all}$ half-short axes of overall ellipse, ($\Delta \ell_{core}$, $\Delta b_{core}$) indicating the center position parameters of the inner ellipse, and ($\Delta \ell_{all}$, $\Delta b_{all}$) the center position parameters of the overall ellipse). The shape parameters in the proper motion space consist of the ellipticity ($e_{pm}$) and the orientation ($q_{pm}$). The values of the ellipticities ($e_{core}$, $e_{all}$ and $e_{pm}$) range between 0 and 1, whereas $q_{core}$ and $q_{all}$ range from $-90\degr$ to $90\degr$, along with $q_{pm}$.

We calculated the errors of the shape parameters by error transfer. The median of relative errors of the inner ellipticities ($e_{core}$, corresponding to the inner shape of the clusters) for the sample clusters is about 4\%, and that of the overall ellipticities ($e_{all}$, corresponding to the overall shape of the clusters) is about 10\%. The median of relative errors of the ellipticity ($e_{pm}$) of the clusters in the proper motion space is about 15\%. We note that several clusters in our sample do not display a reliable solution in the two-dimensional spherical Galactic coordinate system or the proper motion space with our analysis pipeline for the overall ellipticity due to the scattered distribution of its members; thus, the overall shape of these clusters was estimated by visual inspection for it. All shape parameters of our sample and the errors corresponding to these shape parameters are compiled in Table~\ref{table:data1}.

\begin{table*}

  \centering
  \caption{Parameters of 1256 open clusters in this study}

  \label{table:data1}
 % \small
  \tiny
   \resizebox{\textwidth}{!}
{\begin{tabular}{cccccccccccccccccc}
  \hline\noalign{\smallskip}
  \hline\noalign{\smallskip}
  (1) & (2) & (3) & (4) & (5) & (6) & (7) & (8) & (9) & (10) & (11) & (12) & (13) & (14) & (15) & (16) & (17) & (18)\\
  Name & $e_{core}$ & $\sigma_{core}$ & $q_{core}$ & $e_{all}$ & $\sigma_{all}$ & $q_{all}$ & $a_{core}$ & $\sigma a_{core}$ & $c_{core}$ & $\sigma c_{core}$ & $a_{all}$ & $\sigma a_{all}$ & $c_{all}$ & $\sigma c_{all}$ & $e_{pm}$ & $\sigma_{pm}$ & $q_{pm}$ \\
 % \hline\noalign{\smallskip}
  \cline{7-15}
   &  &  & (degree) &  & & \multicolumn{9}{c}{(degree)}                  &  & & (degree) \\
  \hline\noalign{\smallskip}
  ASCC\_10           & 0.182  & 0.037  & -10.040  & 0.264  & 0.107  & -3.115  & 0.497  & 0.014  & 0.406  & 0.014  & 1.635  & 0.140  & 1.203  & 0.140  & 0.158  & 0.189  & 67.060  \\
  ASCC\_101          & 0.354  & 0.027  & 13.013  & 0.432  & 0.194  & 5.701  & 0.406  & 0.009  & 0.262  & 0.009  & 1.392  & 0.235  & 0.791  & 0.235  & 0.161  & 0.180  & -84.501  \\
  ASCC\_105          & 0.197  & 0.016  & -9.999  & 0.075  & 0.080  & -57.399  & 0.677  & 0.008  & 0.544  & 0.008  & 2.373  & 0.139  & 2.196  & 0.139  & 0.128  & 0.155  & 59.116  \\
  ASCC\_107          & 0.194  & 0.036  & 18.895  & 0.155  & 0.071  & 7.634  & 0.196  & 0.005  & 0.158  & 0.005  & 0.555  & 0.030  & 0.469  & 0.030  & 0.163  & 0.332  & 73.558  \\
  ASCC\_108          & 0.418  & 0.015  & -0.084  & 0.219  & 0.073  & -10.958  & 0.570  & 0.007  & 0.332  & 0.007  & 1.855  & 0.106  & 1.449  & 0.106  & 0.107  & 0.067  & 69.118  \\
  ...  &  ...  &  ...  &  ...  &  ...  &  ...  &  ...  &  ...  &  ...  &  ...  &  ...  &  ...  &  ...  &  ...  &  ...  &  ...  &  ...  &  ...  \\

  \hline\noalign{\smallskip}
  \end{tabular}}
  \tablefoot{Column (1) the cluster name (Name), Columns (2) and (5) are the ellipticities ($e_{core}$ and $e_{all}$) of the sample clusters. Columns (3) and (6) are the errors corresponding to the two ellipticities, respectively. Columns (4) and (7) are the orientations corresponding to the two parts in degree ($q_{core}$ and $q_{all}$), respectively. Columns (8) and (10) are half-length and half-short axes of the inner ellipses, respectively. The errors of columns (8) and (10) are displayed in Columns (9) and (11), respectively. Columns (12) and (14) are half-length and half-short axes of the overall ellipses, respectively. The errors of columns (12) and (14) are displayed in Columns (13) and (15), respectively. The ellipticities ($e_{pm}$) in the proper motion space and its error are presented in columns (16) and (17), respectively. Column (18) is the orientation corresponding to column (16). The complete table is available in machine-readable form.}
  \flushleft
\end{table*}

\section{Results}

We present a presentation of the results we obtain from an observational point of view in this section. In our work, we expand the previous sample clusters (265 open clusters) studied in Paper~I to the large sample clusters (1256 open clusters) by a factor of ~5. In the following subsections, we intend to study the morphological evolution of the large sample clusters in the two-dimensional spherical Galactic coordinate system. Based on the same approach with Paper~I, the shape of the sample clusters in proper motion space is simply investigated. The study about the clusters' shape in proper motion space is conducted for the first time, which is the main novelty of our work. We combine the clusters' morphology in the two-dimensional spherical Galactic coordinate system to that in the proper motion space to analyze the morphological evolution trend. At the end of this section, we continue to investigate the degree of stratification of open clusters defined in Paper~I on the basis of the large sample. A glossary of terms involving the newly defined physical parameters is presented in Table~\ref{table:data2}.

\begin{table}[ht]
\caption{Glossary of newly introduced quantities.}
\centering

\label{table:data2}
\resizebox{0.5\textwidth}{!}
{\begin{tabular}{ll}
\hline\noalign{\smallskip}
\hline\noalign{\smallskip}
New Parameters & Name \\
(1) & (2) \\
\hline\noalign{\smallskip}
$d$ & the dislocation of clusters \\

$\Delta a/a_{core}$  &  the relative degree of deformation in the long-axis direction \\

$\Delta c/c_{core}$  &  the relative degree of deformation in the short-axis direction  \\

$|q_{all}$-$q_{core}$$|$$_{valid}$   &  the morphological distortion \\

\hline\noalign{\smallskip}
\end{tabular}}
\tiny
{\tablefoot{Column~1 and 2 represent the symbols for new physical quantities and the name that represents these new quantities, respectively.}}
\flushleft
\end{table}

\begin{figure}[ht]
\centering
\includegraphics[width=65mm]{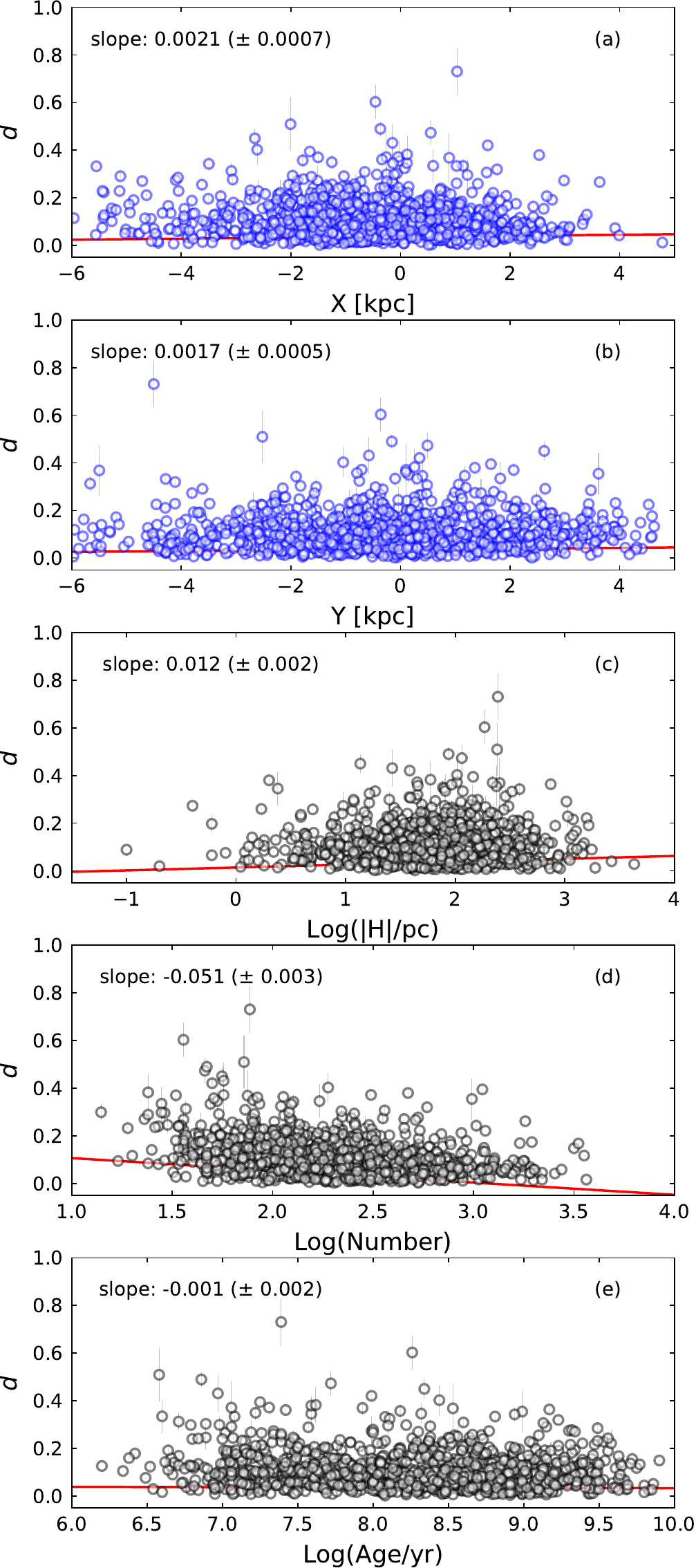}
\caption{
Dislocation between the inner and overall morphology center of the sample cluster in the two-dimensional Galactic reference frame centered on the Sun versus five parameters (e.g., parameters, X, Y, and log(|H|/pc) -- the Cartesian coordinates in a Galactic reference frame centered on the Sun, the number of member stars, and the age of clusters. Each open circle colored (blue and grey) in each panel indicates a cluster, with its gray error bar displaying the dislocation error of each cluster. The corresponding linear fit in each panel is shown as a red line. We note that the coordinate parameters (X, Y, and H), the number of members, and age parameters are obtained from \citet{cant20a}; also, the blue circles in a and b panel only indicate part of the sample clusters that is located within the range from -6~kpc to 5~kpc for both the X and Y-axis. This is due to the fact that the sample clusters outside of this scope are too few to ascertain a reliable fitting result. The grey circles in each rest panel represent entire sample clusters.
}
\label{fig:distance}
\end{figure}

\subsection{Dislocation between the inner and overall morphological center of the sample cluster}

\label{Sec:Dislocation}

We aim to analyze the morphological evolution of open clusters, therefore, it is important to seek or define a new physical parameter to estimate the morphological property of our sample clusters. It is expected that the new physical parameter can better be used to describe the system evolution of clusters and also trace the external environment. To this end, we define $d$ as the dislocation between the inner and overall morphological center of the sample clusters in the two-dimensional spherical Galactic coordinate system. The dislocation, $d,$ is defined following the equation:

\begin{footnotesize}
\begin{equation}
\centering
d = \cfrac{\sqrt{(\Delta \ell_{all} - \Delta \ell_{core})^2 + (\Delta b_{all} - \Delta b_{core})^2}}{a_{all}}
,\end{equation}
\end{footnotesize}

where ($\Delta \ell_{core}$, $\Delta b_{core}$) and ($\Delta \ell_{all}$, $\Delta b_{all}$) denote the center position parameters of the inner and overall ellipse, respectively, with $a_{all}$ representing the half-length axes of the overall ellipse. The dislocation of the sample clusters is possibly caused by external disturbance, such as encounters with giant molecular clouds and spiral arms as well as interactions with the tidal field. Therefore, the dislocation, $d,$ enables thetracing of the complexity of the external environment in the Milk Way. Assuming that the greater $d$ is, the more complex the external environment in which it is located. In addition, the dislocation of the cluster system may be also affected by two-body relaxation and stellar evolution since, at early stages, the cluster consists of separate subclusters \citep{tutu78,cart04,gute08,kuhn14}. Such a structure is slowly changed as a result of massive stellar evolution and gas dissipation, also thereby influenced by two-body relaxation.

Here, we attempt to investigate the relation between the dislocation, $d,$ and the spacial position, as well as with the age parameters and the number of clusters' members. Figure~\ref{fig:distance} plots the dislocation of sample clusters against these parameters. With reference to Figs~\ref{fig:distance}a, b, and c, we find that the dislocation, $d,$ of our sample clusters is related to the X-axis pointing toward the Galactic center and the Y-axis pointing in the direction of Galactic rotation. Meanwhile, it also has a correlation with the Z-axis that is positive toward the Galactic north pole (the absolute height (log(|H|/pc)) from the Galactic disk). This argues possibly that the Galactic disk presents an uneven external environment along the direction of these three axes. As expected, there is a negative correlation between the dislocation and the number of cluster members; sees Figs~\ref{fig:distance}d. The reason for this is that the larger the number of member stars, the stronger the self-gravity of the cluster and the smaller the dislocation of the cluster system is expected to be. Meanwhile, we find from Figs~\ref{fig:distance}e that there is no correlation between the dislocation and the age of clusters. We note that the correlation of Figs~\ref{fig:distance}c is possibly affected by the number of sample clusters in the first half of the horizontal axis, thus, a larger sample is needed to ascertain the behavior.

\subsection{Morphology of the sample clusters in proper motion space}

It is typically accepted that the members of open clusters form almost in the same molecular cloud. Therefore, they share a similar chemical and kinematic property while at the same time presenting a clustered distribution in the proper motion space. However, the shape of the clustered distribution has not been related to more studies in a systematic way. Inspired by Paper~I, in this subsection, we attempt to statistically research the morphology of clusters in the proper motion space. Finally, we try to relate the morphology of the clusters in the proper motion space to the shape of the clusters in the two-dimensional spherical Galactic coordinate system.

\subsubsection{Shape Parameters in the proper motion space}

\label{Sec:Shape Parameters}

The morphology of clusters in the proper motion space is actually a statistic of the magnitude and direction of the proper motion vector of cluster members. If the proper motion vectors of the members of a cluster share roughly the same direction (positive or negative), the shape of the cluster in the proper motion space may be an elliptical morphology where the size of the ellipticity ($e_{pm}$) is possibly dependent to the velocity magnitude of the proper motion of the members in the cluster. The direction along the long axis of the ellipse, orientation ($q_{pm}$), can be viewed as the direction where the clusters' morphology in the two-dimensional spherical Galactic coordinate system is most likely to undergo severe deformation in the future. It is important for the trend of morphological evolution of clusters we study.

\begin{figure}[ht]
\centering
\includegraphics[width=65mm]{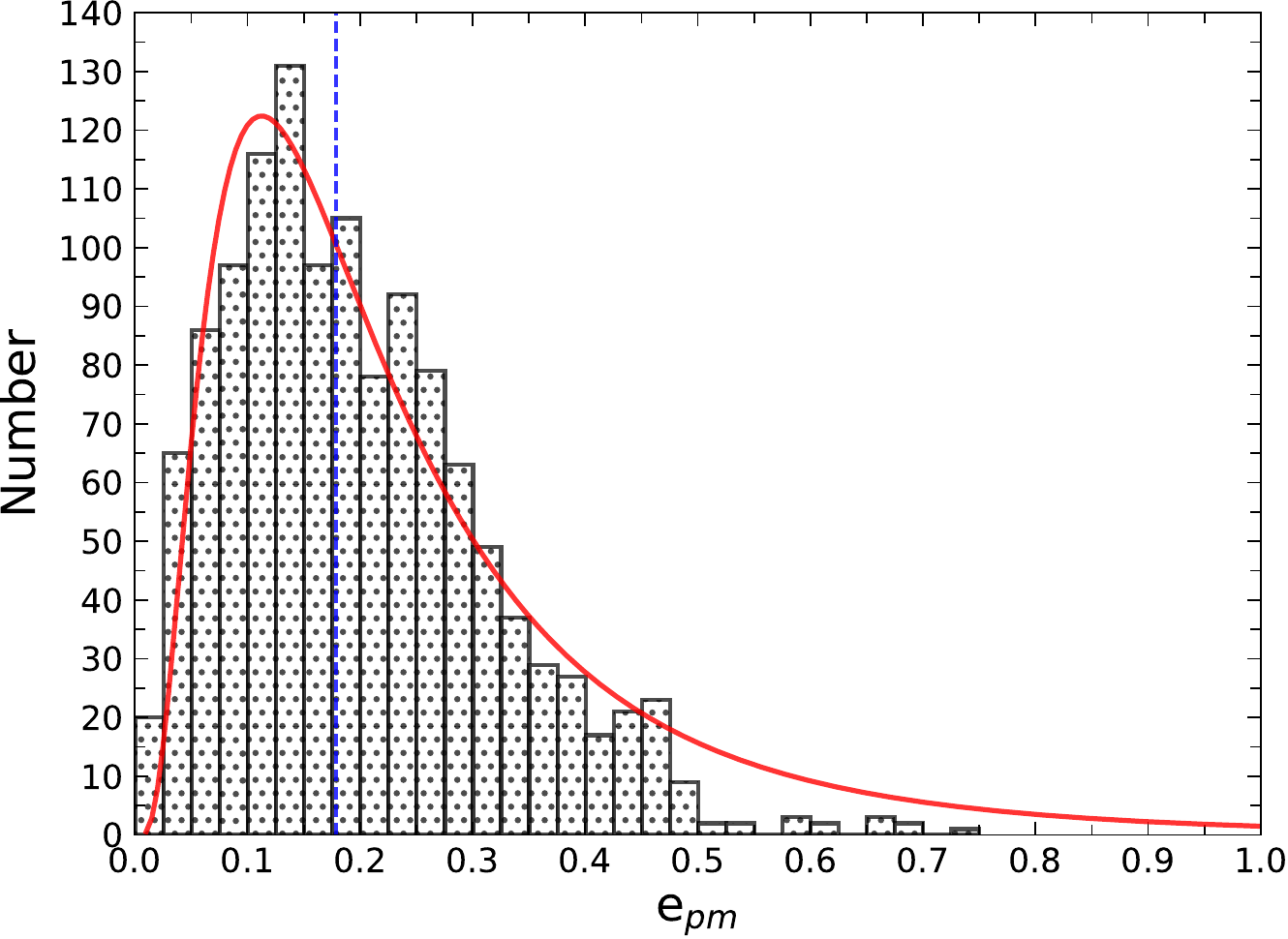}
\caption{
Histogram of the ellipticity ($e_{pm}$) in the proper motion space. The red curve line is the lognormal fitting line to the histogram. The blue dashed line represents the location of the median of the ellipticity ($e_{pm}$).
}
\label{fig:Pm_distrbution}
\end{figure}

To analyze the shape parameters we derived in the proper motion space, we plot the histogram of the ellipticity ($e_{pm}$) of sample clusters in the proper motion space in Figure~\ref{fig:Pm_distrbution}. The histogram in the figure is roughly consistent with the logarithmic normal distribution. The evolution of star clusters is in fact a continuous process of losing members and depleting their stability, which is akin to the wear and tear process of a machine. Such a statistical distribution about this process is theoretically consistent with a log-normal distribution. The same is true for the distribution of the ellipticities in the proper motion space. Therefore, we fit the histogram in Figure~\ref{fig:Pm_distrbution} with the logarithmic normal function. It is clear from Figure~\ref{fig:Pm_distrbution} that the distribution has an indistinct tail, which may indicate that parts of clusters with large degree of ellipticity ($e_{pm}$) have disintegrated and become the field stars. However, It is expected that this tail will not disappear because it will be gradually replenished by clusters in the left of the distribution. The median of the histogram is about 0.18. Half of the sample clusters only present a small ellipticity ($e_{pm}$) in the proper motion space. This may illustrate that most sample clusters still possibly maintain  a stable morphological evolution in the future.

\begin{figure}[ht]
\centering
\includegraphics[width=76mm]{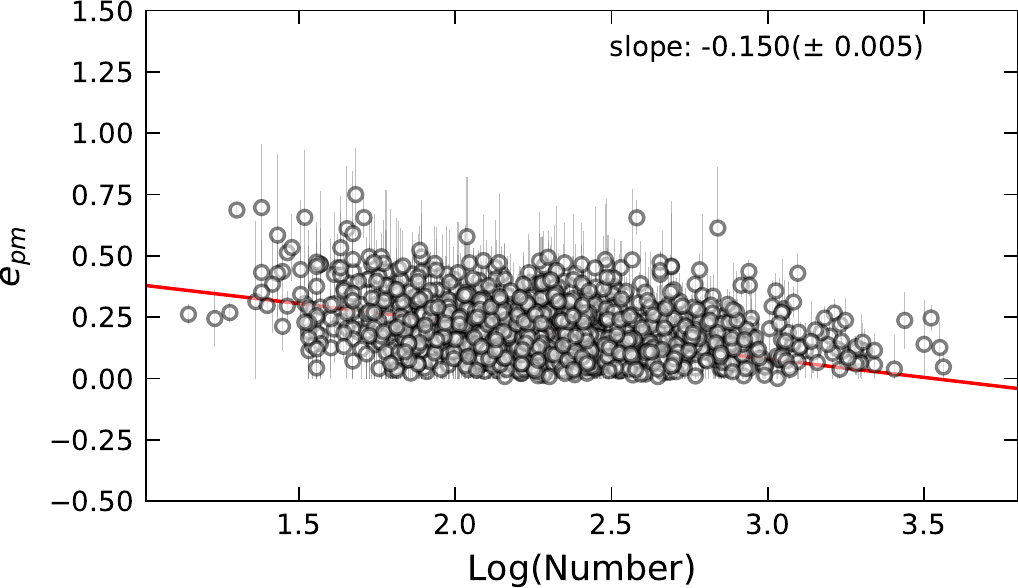}
\caption{
Ellipticities ($e_{pm}$) as a function of the number of clusters' members. The gray bars show the error of the ellipticity. The corresponding linear fit is shown as a red line.
}
\label{fig:Number_pm}
\end{figure}

Furthermore, a significant negative correlation between the ellipticities ($e_{pm}$) in the proper motion space and the numbers of clusters' members is detected for the sample clusters, as shown in Figure~\ref{fig:Number_pm}. This may illustrate that the higher the number of member stars of clusters, the stronger the clusters' gravitational binding to resist the external disturbances, and the more they can keep a smaller ellipticity in the proper motion space. Thus, the clusters with more members should be able to go through a relatively stable shape evolutionary in the proper motion space for a long time in an arcane and complex external environment.

\subsubsection{Combination of the shape parameters in two spaces}

To study the morphological evolution of open clusters in more depth, we use the morphology of clusters in the proper motion space for the first time to investigate the deformation trend of clusters in an exploratory manner.

\begin{figure*}[ht]
\centering
\includegraphics[width=120mm]{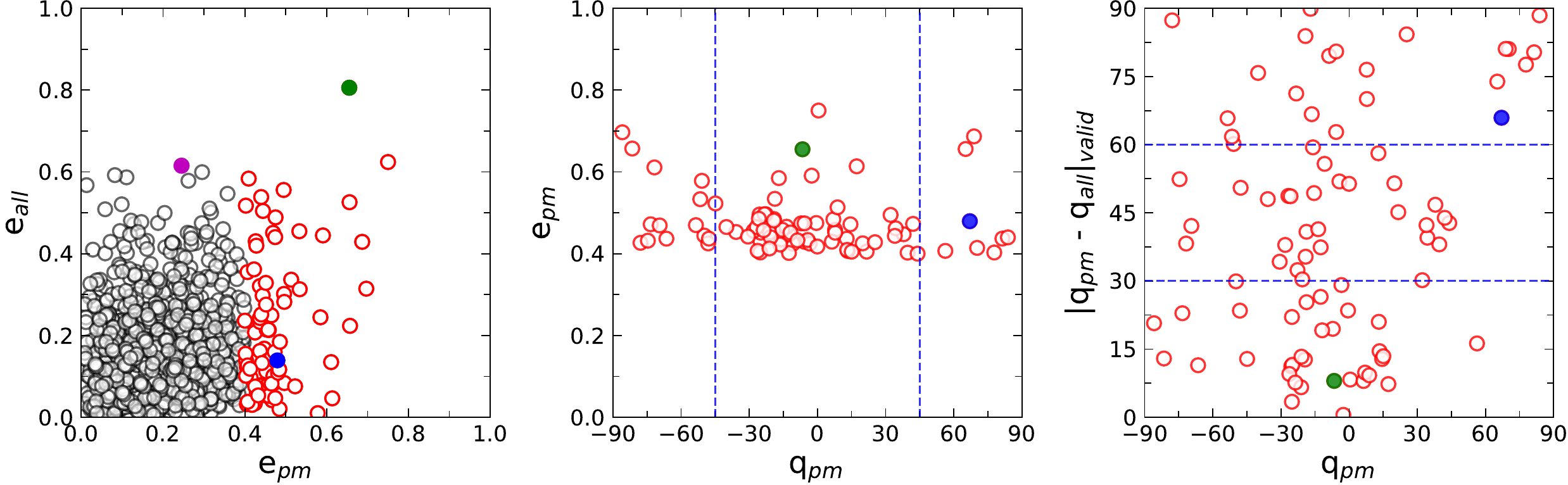}
\caption{
Distribution between the overall ellipticity ($e_{all}$) and the ellipticity ($e_{pm}$) in the proper motion space (left panel), the distribution between the orientations ($q_{pm}$) and the ellipticity ($e_{pm}$) (median panel), and the effective dip angle difference ($|$$q_{pm}$-$q_{all}$$|$$_{valid}$) between the overall orientations ($q_{all}$) and the orientations ($q_{pm}$) in the proper motion space as a function of the orientations ($q_{pm}$) (right panel). Two blue dashed lines in the median panel mark -45$^o$ and 45$^o$ from left to right, respectively. For the right panel, 30$^o$ and 60$^o$ are also marked by two blue dashed lines. The red open circles in each panel belong to the same subsample that is the clusters with $e_{pm}$~$\geq$~0.4, about 7\% of all sample clusters. The blue-filled circle in each panel represents the cluster\ NGC~2264, and the green-filled circle marks the cluster\ Blanco~1. The purple-filled circle in the left panel represents the cluster\ NGC~752.
}
\label{fig:Ellipticity_all_pm_qpm}
\end{figure*}

We plot the overall ellipticity ($e_{all}$) in the two-dimensional spherical Galactic coordinate system against the ellipticity ($e_{pm}$) in the proper motion space to explore the connection that exists between them, as displayed in the left panel of Figure~\ref{fig:Ellipticity_all_pm_qpm}. It can be seen from this panel that there is a slightly positive correlation between the overall ellipticity ($e_{all}$) and the ellipticity ($e_{pm}$), which also has been proved with the Spearman method (see Table~\ref{table:data3}). This might suggest that the overall morphology of sample clusters is possibly dependent on their morphology in the proper motion space. Furthermore, the trend of deformation of the sample clusters' overall morphology may be dependent on its morphology in the proper motion space.

As mentioned above, the orientation ($q_{pm}$) in the proper motion space can be regarded as the direction along which the clusters' morphology in the 2D spherical Galactic coordinate system is most likely to undergo severe deformation in the future. In theory, we can directly study the deformation trend of each sample cluster in the 2D spherical Galactic coordinate system. However, we find that the ellipticities ($e_{pm}$) of most clusters are relatively small. The mean of the ellipticities ($e_{pm}$) in the proper motion space is about 0.2. In fact, there is a false positive for most sample clusters. Because a small ellipticity ($e_{pm}$) may make its orientation ($q_{pm}$) unreliable. It is expected that the clusters with reliable orientation ($q_{pm}$) are screened out for the study of the trend of morphological evolution of clusters. The cluster with $e_{pm}$~$>$~0.2 can be regarded as a fundamentally plausible star cluster that has a reliable orientation ($q_{pm}$) after we analyze the overall data of the sample. However, to improve the confidence of the orientation ($q_{pm}$), we arbitrarily selected the clusters with $e_{pm}$~$\geq$~0.4 as the subsample that is prone to severe deformation in the future.

The median panel of Figure~\ref{fig:Ellipticity_all_pm_qpm} shows the distribution between the orientation ($q_{pm}$) of the subsample clusters in the proper motion space and their ellipticities ($e_{pm}$). We find that the orientation ($q_{pm}$) of the subsample clusters presents an aggregate distribution in the range of -45$\degr$ to 45$\degr$, about 74\% of the subsample. This is not in line with our expectations. But it may suggest that most subsample clusters tend to deform heavily in the direction of the Galactic plane, which is almost consistent with the finding of \citet{mein21}. Most of their sample clusters show a stretched morphology along the direction of the Galactic disk.

One interesting topic to consider is whether the predicted direction ($q_{pm}$) of deformation in a cluster is the same as the direction ($q_{all}$) in which the cluster originally stretched in the two-dimensional spherical Galactic coordinate system. To explore this, we plot the distribution between the orientations ($q_{pm}$) of the subsample clusters and the effective dip angle difference ($|$$q_{pm}$-$q_{all}$$|$$_{valid}$, similar to the definition of $|$$q_{all}$-$q_{core}$$|$$_{valid}$; see Section ~\ref{Sec:Tangential Stratification} in detail) between the overall orientations ($q_{all}$) and the orientations ($q_{pm}$), shown in the right panel in Figure~\ref{fig:Ellipticity_all_pm_qpm}. The predicted direction of deformation for most clusters is not consistent with their originally stretched direction. The division of the three groups in this panel is based on the degree to which  the morphology of clusters in the original stretching direction is deformed. The first group: sample clusters within $|$$q_{pm}$-$q_{all}$$|$$_{valid}$~$\leq$~30$^o$ are defined as the severe deformation clusters. The second group: sample clusters within the range of 30$^o$~$<$~$|$$q_{pm}$-$q_{all}$$|$$_{valid}$~$\leq$~60$^o$ is defined as the moderate deformation clusters. And the last group: sample clusters within the range of 60$^o$~$<$~$|$$q_{pm}$-$q_{all}$$|$$_{valid}$~$\leq$~90$^o$ as the mild deformation clusters. We find that the percentages of the severe deformation clusters, the moderate deformation clusters, and the mild deformation clusters in the entire subsample are 39.5\%, 34.9\%, and 25.6\%, respectively. This may indicate that if nothing else happens, nearly half of the subsample will disintegrate in the future in roughly the original stretching direction. While the mild deformation clusters will mildly deform the shape along the original stretching direction, it is severe for the direction perpendicular to the original stretching direction; whether this deformation trend will prolong or shorten the cluster's lifespan is unknown.

\subsubsection{Prediction of changes in the star cluster morphology}

As mentioned above, the deformation trend of clusters may play an important role in influencing the clusters' lifespan. To investigate the trend of the morphological evolution of clusters in detail, we attempted to analyze some examples. We chose the cluster\ Blanco~1 and\ NGC~2264 from the severe deformation clusters and the mild deformation clusters in the right panel of Figure~\ref{fig:Ellipticity_all_pm_qpm}, respectively.

\begin{figure}[ht]
\centering
\includegraphics[width=76mm]{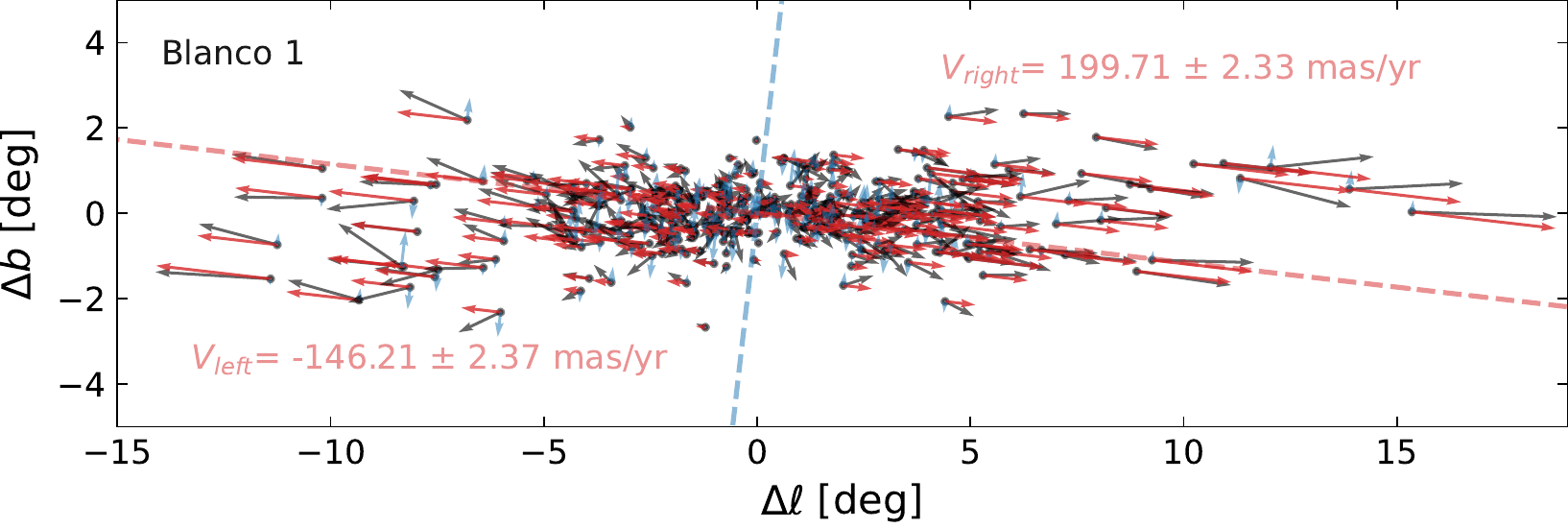}
\caption{Vector distribution of the proper motion of\ Blanco~1. The red dashed line in the figure marks the orientation ($q_{pm}$) of\ Blanco~1, while the blue dashed line marking the direction perpendicular to the orientation. Each gray dot represents its member star, and all gray arrows mark the proper motion vector of the members. All red arrows are the component vector of the proper motion along with the red dashed line, and all blue arrows represent the component vector along the blue dashed line. The length of all arrows represents the size of all vectors.
}
\label{fig:Blanco_1}
\end{figure}

Due to the proximity and the high latitude of\ Blanco~1, it has been widely investigated in past \citep[e.g.,][]{blan49,perr78,deep85,west88,mice99,pill03,pill04,merm08,gonz09,plat11,gaia18,zhan20,mein21}. In our sample,\ Blanco~1 presents a large elliptical structure and has a relative number of members, shown in Figure~\ref{fig:Blanco_1}. The tidal tail of\ Blanco~1 has been discovered by \citet{zhan20}. They also found that\ Blanco~1 is in an early stage of dynamical disintegration. Figure~\ref{fig:Blanco_1} display the vector diagrams of members of\ Blanco~1. We note that those vectors (gray vectors) are decomposed into the predicted direction of deformation (red dashed line) and the direction (blue dashed line) perpendicular to the predicted direction of deformation, respectively. We find that the original stretching direction of\ Blanco~1 is almost the same as its predicted deformation direction, which may indicate that the cluster will continue to deform roughly in the original stretching direction. Significantly, the members of\ Blanco~1 move almost from the center to the left and the right, presenting a pulling situation (see the gray and red arrows in Figure~\ref{fig:Blanco_1}).

The orientation ($q_{pm}$) of the cluster in the proper motion space can be regarded as the most deformed direction of the cluster in the future. The red dashed line in the figure is the most deformed direction which is almost parallel to the original stretching direction. All red arrows given by vector factorization are the vector of proper motion along with the most deformed direction. After counting the sum of the red vectors on both sides of the dashed blue line, respectively, we find that the direction in the left of the blue line is negative, with its value being about -146.21~$\pm$~2.37~mas/yr. This may indicate that the structure in the left of the cluster is going to continue to be stretched, even until it disintegrates and disappears. Similarly, the shape of the structure in the right of the cluster is due to the positive direction and the value of the vector sum being about 199.71~$\pm$~2.33~mas/yr.

\begin{figure}[ht]
\centering
\includegraphics[width=42mm]{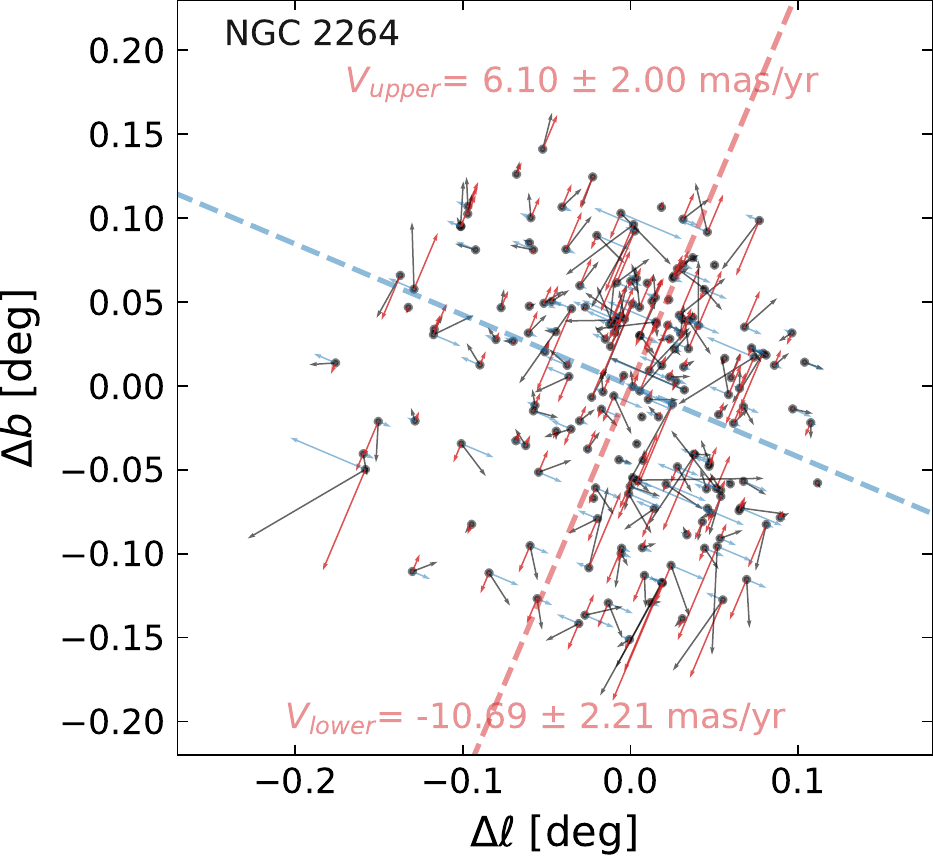}
\caption{
Vector distribution of the proper motion of\ NGC~2264. All symbols are the same with Figure~\ref{fig:Blanco_1}.
}
\label{fig:NGC_2264}
\end{figure}

Furthermore,\ NGC~2264, a young star cluster, also has been widely studied by \citet[][]{venu18,orca19,buck20,nony21}. Figure~\ref{fig:NGC_2264} displays the vector diagrams of members of\ NGC~2264 in the 2D system. It presents a slightly elliptical morphology in the figure. The red dashed line in Figure~\ref{fig:NGC_2264} represents the most deform direction for its original shape in the future. The value of the red vector sum along the most deform direction below the blue dashed line is about -10.69~$\pm$~2.21~mas/yr, while the value of that above the blue dashed line is about 6.10~$\pm$~2.00~mas/yr. This may indicate that\ NGC~2264 may be in a stage of expansion in the 2D space.\ NGC~2264 will deform itself morphology along with the most deform direction in the future if no other events occur.

\begin{figure}[ht]
\centering
\includegraphics[width=45mm]{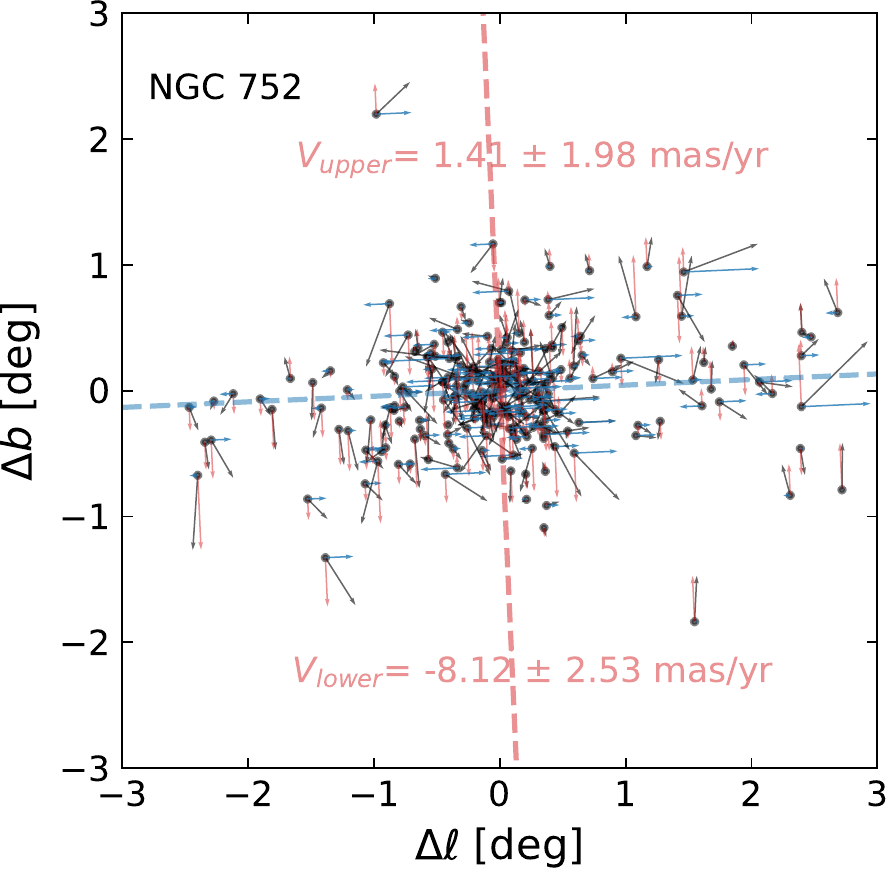}
\caption{
Vector distribution of the proper motion of\ NGC~752. Symbols are the same as in Figure~\ref{fig:Blanco_1}.
}
\label{fig:NGC_752}
\end{figure}

Moreover, there also is a cluster in our sample that resembles\ Blanco~1, namely:\ NGC~752. It is a disintegrating star cluster with a tidal tail \citep{bhat21}. It is not among the clusters with $e_{pm}$~$\geq$~0.4 (see the filled purple circle of Figure~\ref{fig:Ellipticity_all_pm_qpm}). However, it does belong to a plausible star cluster exhibiting a reliable orientation ($q_{pm}$) and its ellipticity in the proper motion space is greater than the mean of the ellipticity ($e_{pm}$) of the sample clusters. Figure~\ref{fig:NGC_752} displays the 2D vector distribution of\ NGC~752. It appears obvious that there is a stretching structure almost along the horizontal direction in the figure. However, unlike\ Blanco~1, it almost presents the most deformed direction perpendicular to the original stretching direction (see the red dashed line in Figure~\ref{fig:NGC_752}). The value of the red vector sum along the most deform direction below the blue dashed line is about -8.12~$\pm$~2.53~mas/yr, while the value of that above the blue dashed line is about 1.41~$\pm$~1.98~mas/yr. Although the value of the vector sum at the top of the figure is small, even less than its error, it will get bigger after a longer period of time. Because  the positive vectors in the bottom can gradually move into the top, with the negative vectors in the top moving into the bottom. This may indicate that\ NGC~752 may be in a slight stage of expansion in the 2D space.\ NGC~752 will deform itself in a morphological sense, along with the most deformed direction, in the future if no other events occur. Whether the original stretching structure of\ NGC~752 continues to exist is unknown and this may need to be further investigated with numerical simulations.

\subsection{Stratification of the layered structure of the sample clusters}

The structure of an open cluster resembles an egg that is made up of a yolk (or core) and an outer layer. Such a structure can be regarded as a layered structure. The term stratification in our work is used to quantify the relationship between the core and the shell of the layered structure. The stratification of the layered structure of open clusters starts was defined and simply studied in Paper~I. The stratifications contain the radial and tangential stratification degree in Paper~I. However, due to the small number of previous sample clusters, the study to analyze the stratification degree of clusters is not enough to derive more interesting laws. In the present work, we still use the shape parameters of our sample clusters to research the stratification degree. The ($\Delta a/a_{core}$) and ($\Delta c/c_{core}$) still are regarded as the radial stratification degree inside the cluster. However, in this paper, we renamed it the "relative degree of deformation." In detail, ($\Delta a/a_{core}$) denotes the relative degree of deformation in the long-axis direction, and ($\Delta c/c_{core}$) denotes the relative degree of deformation in the short-axis direction. Furthermore, $\Delta a$ and $\Delta c$ are equal to $a_{all}-a_{core}$ and $c_{all}-c_{core}$, respectively. However, the original "tangential stratification" ( $q_{all}-q_{core}$)  may be inappropriate after considering the actual situation. First, we renamed it the "degree of morphological distortion," as well as the "morphological distortion." We redefined its formula to be: $|$$q_{all}$-$q_{core}$$|$$_{valid}$ (see Section \ref{Sec:Tangential Stratification} for details).

\subsubsection{Relative degree of deformation}

The relative degree of deformation is a quantification of the cluster layered structure, which can be further divided into the relative degree of deformation in the long-axis direction and that in the short-axis direction. $\Delta a/a_{core}$ and $\Delta c/c_{core}$ represent the relative degree of deformation in the long-axis direction and that in the short-axis direction, respectively. In detail, $\Delta a$ and $\Delta c$ are equal to $a_{all}$-$a_{core}$ and $c_{all}$-$c_{core}$, respectively. The relative degree of deformation of clusters can be viewed as an indicator to estimate the evolutionary degree of open clusters. Assuming that the greater the relative degree of deformation, the greater the degree of evolution of the cluster morphology.

\begin{figure}[ht]
\centering
\includegraphics[angle=0,width=72mm]{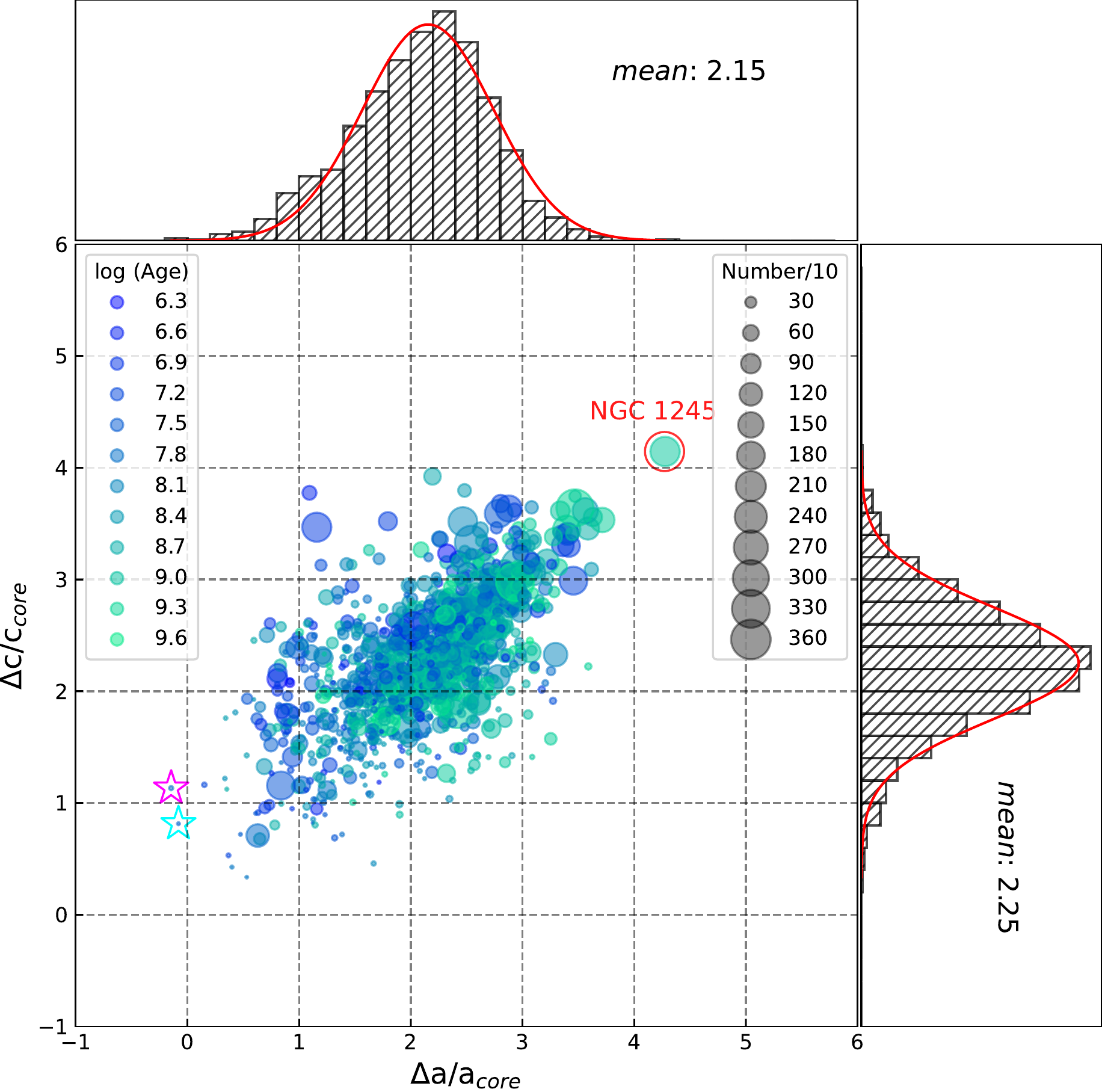}
\caption{
$\Delta c/c_{core}$ distribution function for sample clusters with their $\Delta a/a_{core}$ distribution. The color shade of the solid circles indicates the logarithmic ages (years) of the sample clusters. The size of the symbol is proportional to the number of sample cluster members. The magenta pentagram represents the cluster\ $UBC14$, and the cluster\ $UPK88$ is marked in the cyan pentagram. The dashed lines in the picture are the grid lines for the radial stratification distribution of sample clusters. Histograms of $\Delta c/c_{core}$ and $\Delta a/a_{core}$ of the sample clusters are displayed at the right side and top of the picture, respectively. The red curve lines over the histograms are Gaussian fitting lines.
}
\label{fig:Radial}
\end{figure}

Figure~\ref{fig:Radial} displays the distribution between $\Delta a/a_{core}$ and $\Delta c/c_{core}$ of the sample clusters. For our sample clusters, most clusters locate in the range of the relative degree of deformation from 1 to 3, both in the long-axis and short-axis direction, see the top and right side of Figure~\ref{fig:Radial}. The mean of $\Delta a/a_{core}$ of the sample clusters is 2.15, with the mean of $\Delta c/c_{core}$ being 2.25. It is interesting to note that there are three square areas almost filled by the sample clusters in Figure~\ref{fig:Radial}: one square area, the range of the square area from 1 to 2 both on the x-axis and y-axis direction, is mostly filled by the clusters consisting of a small number of member stars (about 72\% in this square). The clusters in the square area as a percentage of the total sample are about 17\%. The second square area located in the range from 2 to 3 both on the x-axis and y-axis direction and is occupied by most clusters with a large number of members (about 63\% in this square). Meanwhile, all clusters in the second square as a percentage of the whole sample are about 41\%. The last square area in the ranges from 1 to 2 on the x-axis and from 2 to 3 on the y-axis also consists of clusters that are about 17\% of all sample clusters. The square area also contains about 63\% of clusters that consist of a small number of members. This may demonstrate that the starting interval for the relative degree of deformation in the sample clusters with a high number of member stars is about between 2 and 3, while for the sample clusters with a low number of members, it is about between 1 and 2.

In Figure~\ref{fig:Radial}, we find that the age of the sample clusters is possibly related to the relative degree of deformation. Then we adopted $p$, which is the ratio of the relative degree of deformation in the short-axis direction to that in the long-axis direction,  to study the possible correlation between the age of the sample clusters and the relative degree of deformation. The $p$ has been defined in Paper~I. In detail, $p$~$>$~1 shows that the relative degree of deformation in the short-axis direction is greater than that of the long-axis, while the opposite is true for $p$~$\leq$~1. Counting statistics for the $p$, we find that for young sample clusters with log(age/year)~$\leq$~7, in about 76\% of the clusters the relative degree of deformation of the short-axis is greater than that of the long-axis direction ($p$~$>$~1), which is consistent with the finding of Paper~I, while in the clusters with (7~$<$~ log(age/year)~$\leq$~9), about 63\% of the clusters, the relative degree of deformation of the short-axis is greater than that of the long-axis direction. Furthermore, the relative degree of deformation in the short-axis direction is almost equal to that in the long-axis direction for the remaining old clusters. These conclusions seem to indicate the relative degree of deformation of the short-axis of sample clusters decreases with their ages increasing. This conclusion cannot be drawn from Paper~I due to the number of sample clusters.

There is the cluster\ NGC~1245 marked by the red circle in Figure~\ref{fig:Radial}. Noting that its relative degree of deformation is the largest one in our sample cluster, which is greater than 4 both in two directions.\ NGC~1245 is an old (log(age) = 9.08) and populous (Number~=~1991) cluster located in about~448.4~pc perpendicular to the Galactic plane. After checking its ellipticities ($e_{core}$ and $e_{all}$), we find that it owns a circular morphology in core and overall. Although it may be abnormal in shape, a large number of member stars in the\ NGC~1245 can keep itself less elliptical. However, its large relative degree of deformation may indicate that it shows a large degree of evolution, which is almost consistent with the finding of \citet{akma20}. They found that\ NGC~1245 has been in the stage of mass segregation, that is low-mass stars in the core of the cluster are transferred to the cluster's outskirts, while massive stars sink in the core. How long the outer shape consisting of many low-mass sources is altered completely should be focused on in detail.

Curiously two (\ UBC14 marked by in a magenta pentagram and\ UPK88 in a cyan pentagram) of the sample clusters exhibit a negative relative degree of deformation, shown in Figure~\ref{fig:Radial}. We find that their members are few, with a space distribution like field stars. This may challenge their identity for they are newly identified open clusters. While they are young now, they are likely to be disintegrated soon afterward, if they are open clusters. Because the $a_{all}$ is less than the $a_{core}$ in the two clusters, that is to say, the cores of the two clusters are almost non-existent. These also should be further validated and studied.

\subsubsection{Morphological distortion}

\label{Sec:Tangential Stratification}

Similarly to the relative degree of deformation of clusters, the tangential stratification defined in Paper~I ($q_{all}$-$q_{core}$) of clusters is another indicator used to describe the layered structure of open clusters, which can evaluate the distortion of cluster morphology. The greater the value of $q_{all}$-$q_{core}$, the more severe the distortion of the cluster morphology. The distortion effect should be attributed to some different external forces, such as the Galactic tides, the Galactic differential rotation and encounters with giant molecular clouds. The $q_{all}$-$q_{core}$ is simply an expression of the theoretical degree of tangential stratification. It is not suitable to represent the actual tangential stratification degree, so the effective dip angle difference ($|$$q_{all}$-$q_{core}$$|$$_{valid}$) will be used as a new expression of tangential stratification degree. We also rename it the morphological distortion. The morphological distortion ($|$$q_{all}$-$q_{core}$$|$$_{valid}$) is defined following the equation:

\begin{footnotesize}
\begin{equation}
|q_{all}-q_{core}|_{valid}=
\begin{cases}
|q_{all}-q_{core}|& \text{$|q_{all}-q_{core}|$$<$$90^{o}$},\\
180^{o}-|q_{all}-q_{core}|& \text{$|q_{all}-q_{core}|$$\geq$$90^{o}$}.
\end{cases}
\end{equation}
\end{footnotesize}

\begin{figure*}[ht]
\centering
\includegraphics[angle=0,width=180mm]{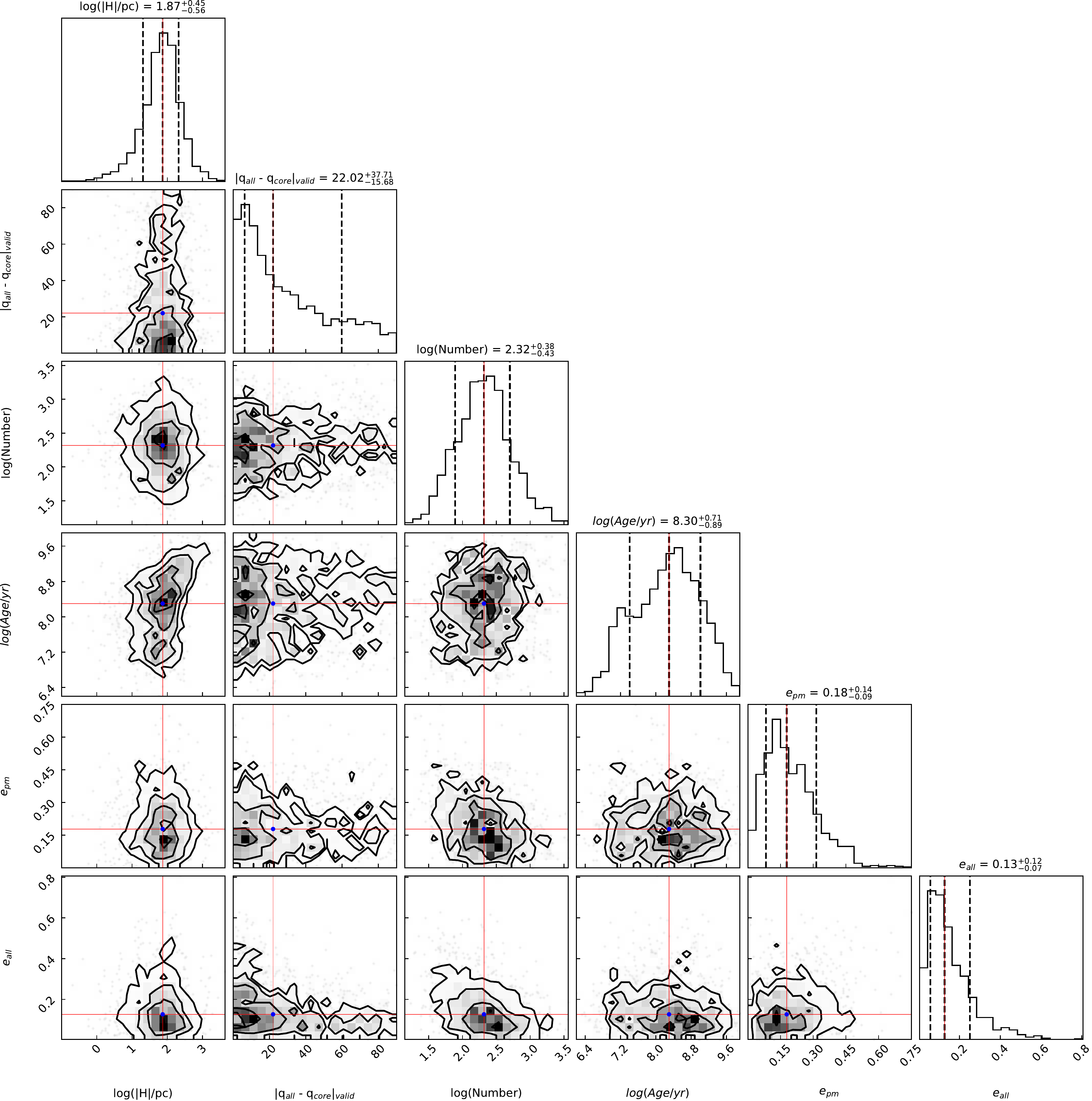}
\caption{
Corner plot for six parameters: age, number of members, absolute height from the plane, overall ellipticity ($e_{all}$), ellipticity ($e_{pm}$) in the proper motion space, and the morphological distortion ($|$$q_{all}$-$q_{core}$$|$$_{valid}$). In the six diagonal panels, we show the probability distribution functions of the variables with the 16th, 50th, and 84th percentiles of the distribution (black dashed lines from left to right). The panels with the maps show the 2D probability for each couple of parameters. The red line in each panel marks the median of each panel. The blue point in each panel is the intersection of two median lines each panel. We note that the parameters H, the number of members, and the age parameter are obtained from \citet{cant20a}.
}
\label{fig:Tangential}
\end{figure*}

We intend to explore whether the morphological distortion might be influenced by other factors such as age, the number of member stars, and the ellipticities in two different spaces. In Figure~\ref{fig:Tangential}, the corner plot containing six parameters is displayed. There are some correlations between some parameters in the corner plot. To verify it, with Spearman correlation estimation we calculate then the correlation coefficients of these parameters and probability for rejecting the null hypothesis that there is no correlation. All correlation coefficients between parameters are listed in Table~\ref{table:data3}. It can be seen from this table that the morphological distortion of clusters is related to their overall ellipticity ($e_{all}$). This correlation is negative, which means probably that the clusters with broad morphological distortion are prone to presenting a circular external shape. The physical mechanism behind this remains to be explored. Even though only one correlation is detected in the morphological distortion, some correlations between other parameters are uncovered. This is a feature that is also interesting in the context of our work.

\begin{table}[ht]\tiny
\caption{Correlation data}
\centering

\label{table:data3}
\begin{tabular}{c c c}
\hline\noalign{\smallskip}
\hline\noalign{\smallskip}
Variables & Spearman & $p$-value \\
(1) & (2) & (3) \\
\hline\noalign{\smallskip}
log(Age/yr) versus $e_{all}$  & -0.11 & 1.40~$\times$~$10^{-4}$ \\

log(Number) versus $e_{all}$  & -0.38  & 2.12~$\times$~$10^{-43}$ \\

log(Number) versus $e_{pm}$   & -0.29  & 4.13~$\times$~$10^{-26}$ \\

$|$$q_{all}$-$q_{core}$$|$$_{valid}$ versus $e_{all}$  & -0.30 & 7.31~$\times$~$10^{-27}$ \\

$e_{pm}$ versus $e_{all}$  &  0.16 & 8.39~$\times$~$10^{-9}$ \\

log(Age/yr) versus log(Number)   & 0.12  & 3.50~$\times$~$10^{-5}$ \\

log($|$H$|$/pc) versus log(Age/yr)   & 0.40  & 1.15~$\times$~$10^{-48}$ \\

\hline\noalign{\smallskip}
\end{tabular}
\tablefoot{Column~1,2, and 3 represent the name of variables, Spearmen's $\rho$ correlation coefficient and probability for rejecting the null hypothesis that there is no correlation, respectively.}
\flushleft
\end{table}

The negative correlation of the overall ellipticity ($e_{all}$) with the age of clusters and the number of members is consistent with the finding detected in Paper~I. In Section \ref{Sec:Shape Parameters}, the negative correlation between the number of member stars and the ellipticity in the proper motion space is found. And we again confirm this correlation in the present section. In addition, we find also that the ellipticity ($e_{pm}$) in the proper motion space is related to the overall ellipticity ($e_{all}$) of sample clusters. The age of clusters has a positive correlation with the number of members, also with the absolute height from the Galactic disk. The latter is roughly consistent with the result of \citet{chen04}; for more, see Fig~4 of their paper.

\begin{figure}[ht]
\centering
\includegraphics[angle=0,width=82mm]{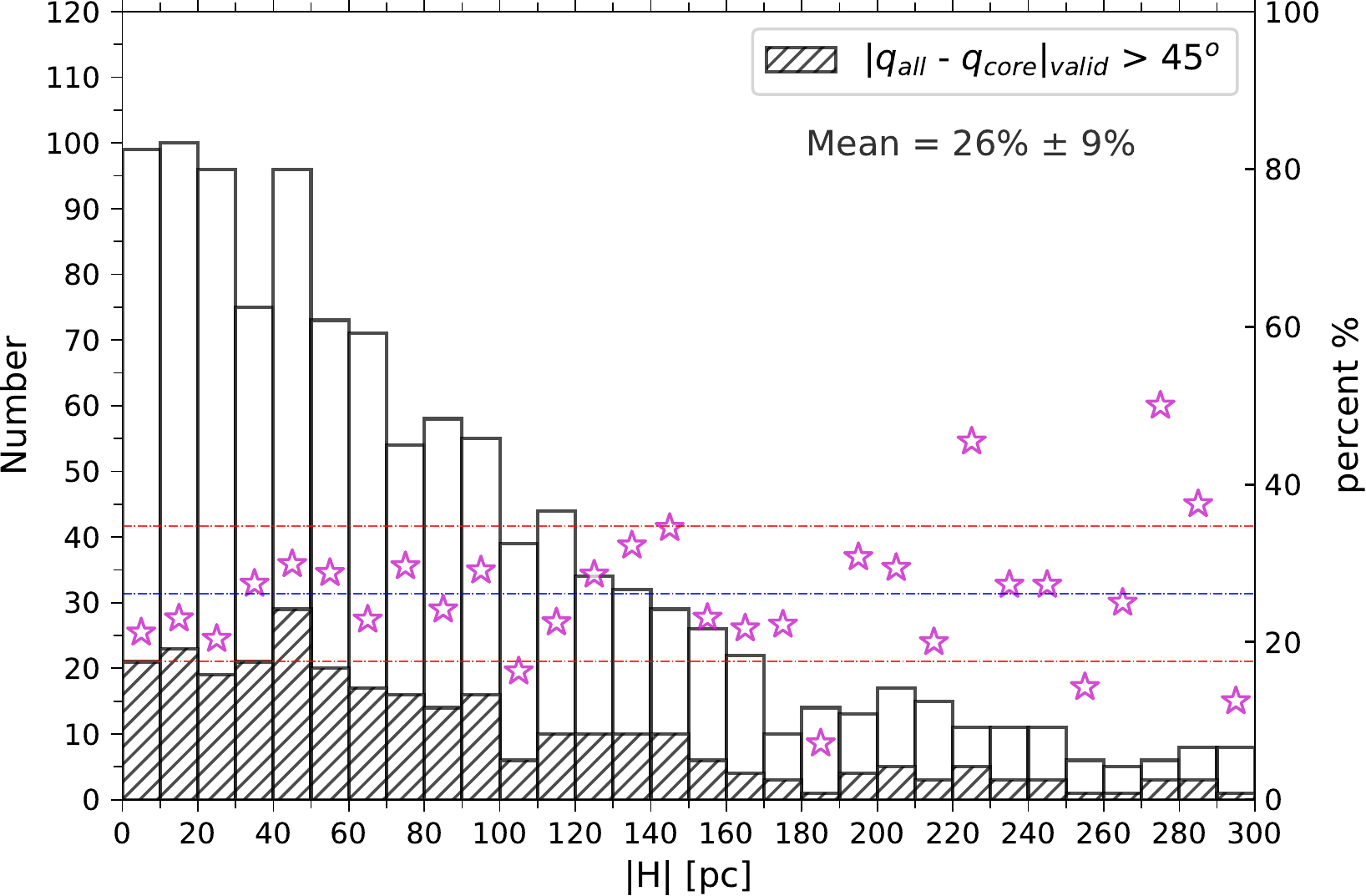}
\caption{
  Histogram of the sample clusters with $|$H$|$~$\leq$~300~pc and the distribution of the proportion of the sample clusters with the effective dip angle difference greater than 45$^o$ from each interval to the total clusters in each interval. The empty histogram represents the total clusters in each interval, while the shadow histogram the sample clusters with $|$$q_{all}$-$q_{core}$$|$$_{valid}$~$>$~45$^o$ in each interval. The open purple pentagram in the panel represents the proportion of the sample clusters with the effective dip angle difference greater than 45$^o$ from each interval to the total clusters in each interval. The blue dashed line marks the mean of these proportions and the red dashed line represents its error.
}
\label{fig:Group}
\end{figure}

While the correlation between the morphological distortion and the absolute height from the Galactic disk was not found, the relation of the dislocation of clusters with the absolute height was detected, as described in Section \ref{Sec:Dislocation}. Therefore, we again investigated the possible relation between the morphological distortion and the absolute height from the Galactic disk. First, the sample clusters with $|$H$|$~$\leq$~300~pc were grouped into 30 groups according to the step size of 10~pc, as shown in Figure~\ref{fig:Group}. Then the clusters whose effective dip difference is greater than 45$^o$ in each group were counted. Here, we regard the cluster with $|$$q_{all}$-$q_{core}$$|$$_{valid}$~$>$~45$^o$ as a severely distorted cluster. Interestingly, we find that on average, the severely distorted clusters in each group account for about 26\%~$\pm$~9\% of each group. This may mean that there is a uniform environment of external forces within |H|~$\leq$~300~pc if the sample completeness of each group is not considered. This finding still needs to be verified.

\section{Summary}

We present a novel view of the morphological evolution of sample clusters. With the approach employed in Paper~I, we expand our previous sample clusters to 1256 clusters, then we delve into  the morphological evolution of the sample clusters in the two-dimensional spherical Galactic coordinate system by their shape in the proper motion space, make the relevant predictions, and obtain some new and impressive conclusions. The conclusions are as follows:

%\par

1. We find that the dislocation, $d,$ of clusters is related to the X-axis pointing toward the Galactic center and Y-axis pointing in the direction of Galactic rotation. Meanwhile, this is true for the Z-axis that is positive toward the Galactic north pole (the absolute height (log(|H|/pc)) from the Galactic disk). This argues possibly that the Galactic disk presents an uneven external environment along the direction of these three axes. There is a negative correlation between the dislocation and the number of clusters' members. In addition, no correlation exists between the dislocation and the age of clusters.

2. The ellipticity ($e_{pm}$) exhibits a logarithmic normal distribution and presents an indistinct tail, which may indicate that the clusters with large ellipticity ($e_{pm}$) has disintegrated and has become the field stars. Half of the sample clusters only present a small ellipticity ($e_{pm}$) in the proper motion space, which may suggest that these clusters still keep a stable morphological evolution in the future. A significant negative correlation between the ellipticity ($e_{pm}$) and the number of the member stars is detected for the sample clusters, which indicates probably that the more the number of member stars of the clusters, the stronger the cluster's gravitational binding to resist the external disturbances and the more it can keep a smaller ellipticity ($e_{pm}$) in the proper motion space. There is a slightly positive correlation between the overall ellipticity ($e_{all}$) and the ellipticity ($e_{pm}$), which indicates possibly that the overall morphology of sample clusters is possibly dependent on their morphology in the proper motion space. We find that the orientation ($q_{pm}$) of the sample clusters with $e_{pm}$~$\geq$~0.4 presents an aggregate distribution in the range of -45$\degr$ to 45$\degr$, about 74\% of the subsample, which illustrates possibly that most subsample clusters tend to deform heavily in the direction of the Galactic plane. For sample clusters with $e_{pm}$~$\geq$~0.4, the predicted direction of deformation for most clusters is not consistent with their originally stretched direction. If nothing else happens, the severe deformation clusters will end up disintegrating in the future in roughly the original stretching direction.

3. The orientation ($q_{pm}$) of sample clusters in the proper motion space can better describe the most deform direction of the sample clusters. The member stars of Blanco~1 move almost from the center to the left and the right, presenting a pulling situation.\ NGC~2264 may be in a stage of expansion in the two-dimensional space. Similarly,\ NGC~752 also may be in a slight stage of expansion in the two-dimensional space and will deform itself morphology along the direction perpendicular to the original stretching direction in the future, if no other events occur.

4. The relative degree of deformation of the sample cluster's interior is almost less than 4. Besides, the mean of $\Delta a/a_{core}$ of the sample clusters is 2.15, and the mean of $\Delta c/c_{core}$ 2.25. The starting interval for the relative degree of deformation in the sample clusters with a high number of member stars is greater than the sample clusters with a low number of members. The relative degree of deformation of the short-axis of sample clusters decreases with their ages increasing. By analyzing the relative degree of deformation, it suggests that\ NGC~1245 may have a large degree of evolution. Moreover,\ UBC14 and\ UBC88 both exhibit a negative relative degree of deformation, which means possibly there is no core in their cluster center. So, their cluster identity still needs to be verified.

5. The morphological distortion of clusters is related to their overall ellipticity ($e_{all}$). The negative correlation of the overall ellipticity ($e_{all}$) with the age of clusters and the negative correlation of the $e_{all}$ with the number of members both are consistent with the finding detected in Paper~I. A negative correlation between the number of member stars and the ellipticity in the proper motion space is found. We find also that the ellipticity ($e_{pm}$) in the proper motion space is related to the overall ellipticity ($e_{all}$) of sample clusters. The age of clusters features a positive correlation with the number of members, also with the absolute height from the Galactic disk. On average, the severely distorted clusters in each group account for about 26\%~$\pm$~9\%, which may suggest that there is a uniform environment of external forces at the range of $|$H$|$~$\leq$~300~pc if the sample completeness of each group is not considered.

In our work, we study the morphological evolution of clusters by combining the morphology of the cluster's proper motion space with the morphology of the cluster's real space. This approach is useful for improving our understanding of the morphological evolution of clusters.

\begin{acknowledgements}
This work is supported by the National Natural Science Foundation of China under grants U2031209 and U2031204, the Youth Innovation Promotion Association CAS (grant No.2018080), and the 2017 Heaven Lake Hundred-Talent Program of Xinjiang Uygur Autonomous Region of China. The authors thank the reviewer for the very helpful comments and suggestions. We would also like to thank Ms. Chunli Feng for touching up the language of the article. This study made use of the Gaia DR2, operated at the European Space Agency (ESA) space mission (Gaia). The Gaia archive website is \url{https://archives.esac.esa.int/gaia/}.

Software: Astropy \citep{astr13,astr18}, Scipy \citep{mill11}, TOPCAT \citep{tayl05}.

\end{acknowledgements}


\begin{thebibliography}{}

\bibitem[Astropy Collaboration et al.(2013)]{astr13} Astropy Collaboration, Robitaille, T.~P., Tollerud, E.~J., et al.\ 2013, \aap, 558, A33. doi:10.1051/0004-6361/201322068

\bibitem[Astropy Collaboration et al.(2018)]{astr18} Astropy Collaboration, Price-Whelan, A.~M., Sip{\H{o}}cz, B.~M., et al.\ 2018, \aj, 156, 123. doi:10.3847/1538-3881/aabc4f

\bibitem[Bhattacharya et al.(2021)]{bhat21} Bhattacharya, S., Agarwal, M., Rao, K.~K., et al.\ 2021, \mnras. doi:10.1093/mnras/stab1404

\bibitem[Blanco(1949)]{blan49} Blanco, V.~M.\ 1949, \pasp, 61, 183. doi:10.1086/126171

\bibitem[Buckner et al.(2020)]{buck20} Buckner, A.~S.~M., Khorrami, Z., Gonz{\'a}lez, M., et al.\ 2020, \aap, 636, A80. doi:10.1051/0004-6361/201936935
\bibitem[{\c{C}}akmak et al.(2020)]{akma20} {\c{C}}akmak, H., G{\"u}ne{\c{s}}, O., \& Karata{\c{s}}, Y.\ 2020, arXiv:2012.15587


\bibitem[Cantat-Gaudin et al.(2018)]{cant18} Cantat-Gaudin, T., Jordi, C., Vallenari, A., et al.\ 2018, \aap, 618, A93.
    doi:10.1051/0004-6361/201833476
\bibitem[Castro-Ginard et al.(2019)]{cast19} Castro-Ginard, A., Jordi, C., Luri, X., et al.\ 2019, \aap, 627, A35. doi:10.1051/0004-6361/201935531
\bibitem[Castro-Ginard et al.(2020)]{cast20} Castro-Ginard, A., Jordi, C., Luri, X., et al.\ 2020, \aap, 635, A45. doi:10.1051/0004-6361/201937386
\bibitem[Cantat-Gaudin \& Anders(2020)]{cant20a} Cantat-Gaudin, T. \& Anders, F.\ 2020, \aap, 633, A99. doi:10.1051/0004-6361/201936691
\bibitem[Cantat-Gaudin et al.(2020)]{cant20b} Cantat-Gaudin, T., Anders, F., Castro-Ginard, A., et al.\ 2020, \aap, 640, A1.
    doi:10.1051/0004-6361/202038192

\bibitem[Cartwright \& Whitworth(2004)]{cart04} Cartwright, A. \& Whitworth, A.~P.\ 2004, \mnras, 348, 589. doi:10.1111/j.1365-2966.2004.07360.x

\bibitem[Chen et al.(2004)]{chen04} Chen, W.~P., Chen, C.~W., \& Shu, C.~G.\ 2004, \aj, 128, 2306.
    doi:10.1086/424855

\bibitem[de Epstein \& Epstein(1985)]{deep85} de Epstein, A.~E.~A. \& Epstein, I.\ 1985, \aj, 90, 1211. doi:10.1086/113828

\bibitem[Fall et al.(2005)]{fall05} Fall, S.~M., Chandar, R., \& Whitmore, B.~C.\ 2005, \apjl, 631, L133.doi:10.1086/496878

\bibitem[Gaia Collaboration et al.(2018)]{gaia18} Gaia Collaboration, Babusiaux, C., van Leeuwen, F., et al.\ 2018, \aap, 616, A10. doi:10.1051/0004-6361/201832843
\bibitem[Gutermuth et al.(2008)]{gute08} Gutermuth, R.~A., Myers, P.~C., Megeath, S.~T., et al.\ 2008, \apj, 674, 336. doi:10.1086/524722

\bibitem[Gonz{\'a}lez \& Levato(2009)]{gonz09} Gonz{\'a}lez, J.~F. \& Levato, H.\ 2009, \aap, 507, 541. doi:10.1051/0004-6361/200912772

\bibitem[Hu et al.(2021)]{hu21} Hu, Q., Zhang, Y., Esamdin, A., et al.\ 2021, \apj, 912, 5. doi:10.3847/1538-4357/abec3e

\bibitem[Kholopov(1969)]{khol69} Kholopov, P.~N.\ 1969, \sovast, 12, 625

\bibitem[Krone-Martins \& Moitinho(2014)]{kron14} Krone-Martins, A. \& Moitinho, A.\ 2014, \aap, 561, A57.
    doi:10.1051/0004-6361/201321143

\bibitem[Krumholz et al.(2019)]{krum19} Krumholz, M.~R., McKee, C.~F., \& Bland-Hawthorn, J.\ 2019, \araa, 57, 227. doi:10.1146/annurev-astro-091918-104430

\bibitem[Kuhn et al.(2014)]{kuhn14} Kuhn, M.~A., Feigelson, E.~D., Getman, K.~V., et al.\ 2014, \apj, 787, 107. doi:10.1088/0004-637X/787/2/107

\bibitem[Meingast et al.(2021)]{mein21} Meingast, S., Alves, J., \& Rottensteiner, A.\ 2021, \aap, 645, A84. doi:10.1051/0004-6361/202038610

\bibitem[Mermilliod et al.(2008)]{merm08} Mermilliod, J.-C., Platais, I., James, D.~J., et al.\ 2008, \aap, 485, 95. doi:10.1051/0004-6361:20079072

\bibitem[Micela et al.(1999)]{mice99} Micela, G., Sciortino, S., Favata, F., et al.\ 1999, \aap, 344, 83
\bibitem[Millman \& Aivazis(2011)]{mill11} Millman, K.~J. \& Aivazis, M.\ 2011, Computing in Science and Engineering, 13, 9. doi:10.1109/MCSE.2011.36
\bibitem[Nony et al.(2021)]{nony21} Nony, T., Robitaille, J.-F., Motte, F., et al.\ 2021, \aap, 645, A94. doi:10.1051/0004-6361/202039353

\bibitem[Orcajo et al.(2019)]{orca19} Orcajo, S., Cieza, L.~A., Gamen, R., et al.\ 2019, \mnras, 487, 2937. doi:10.1093/mnras/stz1452

\bibitem[Pang et al.(2021)]{pang21} Pang, X., Li, Y., Yu, Z., et al.\ 2021, \apj, 912, 162. doi:10.3847/1538-4357/abeaac
\bibitem[Perry et al.(1978)]{perr78} Perry, C.~L., Walter, D.~K., \& Crawford, D.~L.\ 1978, \pasp, 90, 81. doi:10.1086/130282

\bibitem[Pillitteri et al.(2003)]{pill03} Pillitteri, I., Micela, G., Sciortino, S., et al.\ 2003, \aap, 399, 919. doi:10.1051/0004-6361:20021818
\bibitem[Pillitteri et al.(2004)]{pill04} Pillitteri, I., Micela, G., Sciortino, S., et al.\ 2004, \aap, 421, 175. doi:10.1051/0004-6361:20035869

\bibitem[Piskunov et al.(2008)]{pisk08} Piskunov, A.~E., Schilbach, E., Kharchenko, N.~V., et al.\ 2008, \aap, 477, 165.
    doi:10.1051/0004-6361:20078525
\bibitem[Platais et al.(2011)]{plat11} Platais, I., Girard, T.~M., Vieira, K., et al.\ 2011, \mnras, 413, 1024. doi:10.1111/j.1365-2966.2011.18194.x

\bibitem[Raboud \& Mermilliod(1998b)]{rabo98b} Raboud, D. \& Mermilliod, J.-C.\ 1998, \aap, 333, 897

\bibitem[Sampedro et al.(2017)]{samp17} Sampedro, L., Dias, W.~S., Alfaro, E.~J., et al.\ 2017, \mnras, 470, 3937. doi:10.1093/mnras/stx1485

\bibitem[Taylor(2005)]{tayl05} Taylor, M.~B.\ 2005, Astronomical Data Analysis Software and Systems XIV, 347, 29

\bibitem[Tutukov(1978)]{tutu78} Tutukov, A.~V.\ 1978, \aap, 70, 57

\bibitem[Venuti et al.(2018)]{venu18} Venuti, L., Prisinzano, L., Sacco, G.~G., et al.\ 2018, \aap, 609, A10. doi:10.1051/0004-6361/201731103

\bibitem[Westerlund et al.(1988)]{west88} Westerlund, B.~E., Garnier, R., Lundgren, K., et al.\ 1988, \aaps, 76, 101

\bibitem[Wielen(1974)]{wiel74} Wielen, R.\ 1974, Stars and the Milky Way System, 169

\bibitem[Zhai et al.(2017)]{zhai17} Zhai, M., Abt, H., Zhao, G., et al.\ 2017, \aj, 153, 57.
    doi:10.3847/1538-3881/153/2/57
\bibitem[Zhang et al.(2020)]{zhan20} Zhang, Y., Tang, S.-Y., Chen, W.~P., et al.\ 2020, \apj, 889, 99.
    doi:10.3847/1538-4357/ab63d4
\end{thebibliography}
\end{document}